%% file: Giancarlo_et_Al.tex
\pdfoutput=1
\documentclass{article}

\usepackage[utf8]{inputenc} 
\usepackage{amsmath,amsthm}
\usepackage{graphicx}
\usepackage{epsfig}
\usepackage{color}
\usepackage{calc}
\usepackage{array}
\usepackage{xspace}
\usepackage{caption}
\usepackage{booktabs}
\usepackage[affil-it]{authblk}

\newcommand{\ignore}[1]{}

\newcommand{\MR}{MapReduce\xspace}
\newcommand{\KV}[2]{\emph{$\langle$${#1}, {#2}$$\rangle$}\xspace}
\newcommand{\FK}{FastKmer\xspace}

\newcommand{\filesystem}{file system\xspace}
\newcommand{\datalocal}{data local\xspace}

\newcommand{\kmer}{$k$-mer\xspace}
\newcommand{\kmers}{$k$-mers\xspace}

\newcommand{\ie}{i.e.,\xspace}

\begin{document}




\title{Analyzing Big Datasets of Genomic Sequences: 
Fast and Scalable Collection of $k$-mer Statistics}

\author{
Umberto Ferraro Petrillo
\thanks{Electronic address: \texttt{umberto.ferraro@uniroma1.it.it}}}
\affil{Dipartimento di Scienze Statistiche, Universit\`{a} di Roma ``La Sapienza'', Italy}

\author{Mara Sorella
\thanks{Electronic address: \texttt{sorella@diag.uniroma1.it}}}
\affil{Dipartimento di Ingegneria Informatica, Automatica e Gestionale, Universit\`{a} di Roma ``La Sapienza'', Italy}

\author{Giuseppe Cattaneo
\thanks{Electronic address: \texttt{cattaneo@unisa.it}}}
\affil{Dipartimento di Informatica, Universit\`a degli studi di Salerno, Italy}

\author{Raffaele Giancarlo
\thanks{Electronic address: \texttt{raffaele.giancarlo@unipa.it}}}

\author{Simona Rombo
\thanks{Electronic address: \texttt{simona.rombo@unipa.it}}}
\affil{Dipartimento  di Matematica ed Informatica, Universit\`{a} di Palermo, Italy}

%
%
%
%

%
%
%
%

%
%
%
%
\date{}
\maketitle
\begin{abstract}
	{\bf Background} Distributed approaches based on the 
map-reduce programming paradigm have started to be proposed in the 
bioinformatics domain, due to the large amount of data produced by the next-generation sequencing techniques. However, the use of map-reduce and related 
Big Data technologies and frameworks (e.g., Apache Hadoop and Spark) does not 
necessarily produce satisfactory results, in terms of both efficiency and 
effectiveness.

{\bf Results} We discuss how the development of distributed and Big Data 
management technologies has affected the analysis of large datasets of 
biological sequences. Moreover, we show how the choice of different 
parameter configurations and the careful engineering of the software with 
respect to the specific framework under consideration may be crucial in order 
to achieve good performance, especially on very large amounts of data. We choose 
\kmers counting as a case study for our analysis, and Spark as the 
framework to implement \FK, a novel approach for the extraction of \kmer 
statistics from large collection of biological sequences, with 
arbitrary values of $k$. One of the most relevant contributions of \FK
 is the introduction of a 
module for balancing the statistics aggregation workload over the nodes of a 
computing cluster, in order to overcome data skew while allowing for a fully 
exploitation 
of the underlying distributed architecture. We also present the results of a 
comparative experimental analysis showing that our approach is currently the 
fastest among the ones based on Big Data technologies, while exhibiting a very 
good scalability.

{\bf Conclusion.} We provide evidence that the usage of 
technologies such as Hadoop or Spark 
for the analysis of big datasets of biological sequences is productive only if 
the architectural details and the peculiar aspects of the considered framework 
are carefully taken into account for the algorithm design and implementation.

\end{abstract}

%
%
%

%
%
%
%
%
%


%
%
%

\input{sec_introduction.tex}
\input{sec_state_of_art.tex}

\input{sec_approach.tex}
\input{sec_refined_approach.tex}

\input{sec_conclusions.tex}

\section*{Acknowledgments}
\subsection*{Funding}
\noindent INdAM - GNCS Project 2017 ``Algoritmi e tecniche efficienti per 
l'organizzazione, la gestione e l'analisi di Big Data in ambito biologico'' and 
INdAM-GNCS Project 2018 ``Elaborazione ed analisi di Big Data modellati come 
grafi in vari contesti applicativi'' to G. Cattaneo, U. Ferraro 
Petrillo, R. Giancarlo and S. E. Rombo.
Cloud computing resources used for the experiments described in this 
paper were provided by a  Microsoft Azure Research award.

\subsection*{Contributions}
All authors contributed to design the research contained in this paper. In 
particular, M. S. and U. F. P. designed the algorithms and performed the 
experimentation. All authors contributed to the writing of the paper. All 
authors have read and approved the manuscript. 
\bibliographystyle{bmc-mathphys}

\end{document}

%% file: sec_introduction.tex

\section{Introduction}

With the rapid growth of biological sequence datasets and the 
evolution of the sequencing technologies, many algorithms and software commonly 
used for the 
analysis of biological sequences are becoming obsolete. For this reason, 
computational approaches based on frameworks for big data processing 
started to be proposed 
in order to deal with problems involving large amounts of biological data 
\cite{fastdoop,CattaneoSUPE2017,Cattaneo2017,Ferraro2017,MetaSpark}. 
Unfortunately, the fundamental domain of alignment-free linguistic and 
informational analysis of genomic and proteomic  sequences, e.g., 
\cite{benoit16,giancarloCompression09,Giancarlo2015, 
LoBosco_LNCS_CIBB_2015,nordstrom13,Pinello11,Giancarlo2015,utro2016},  has 
received yet few attention in the big data context.  In this respect, an 
important task that is at 
the hearth of this domain is the collection of $k$-mer statistics, i.e., how 
many times each sequence of length $k$ over a finite alphabet appears in a set 
of  biological  sequences, at a genomic scale. Once that such information is 
available, one can use it to compute many  informational and linguistic indices
\ignore{e.g.,} \cite{giancarloCompression09,giancarlo2014compressive}, as well as de 
Bruijn graph assemblers, and error/repeat detection systems. 

Recently, the software tool KCH \cite{KCH} has been proposed for the 
linguistic and informational analysis 
of biological sequences based on Hadoop~\cite{HADOOP_BOOK} and 
\MR~\cite{MapReduce}. It
allows for an efficient collection and analysis of 
\kmers from a collection of genomic sequences. KCH has been the first tool 
showing that big data technologies can be superior to highly-optimized shared 
memory multi processor approaches, even when considering mid-size problem 
instances. This latter methodological contribution, combined with results 
in~\cite{Siretskiy2015}, gives ground-breaking experimental evidence that big 
data technologies can be extremely pervasive for an effective solution of a 
broad spectrum of computational problems in the Life Sciences, going from basic 
primitives to full-fledged analysis and storage pipelines.  
However, in quantitative terms, that is only a first step towards the 
acquisition of  full knowledge of how big data technologies can affect 
Computational Biology and Bioinformatics.

In this manuscript we first discuss how the introduction of new, distributed, 
technologies for big data processing is affecting the biological 
domain. Then, we focus on the problem of \kmer counting as a case study, and 
in particular we present a distributed approach for \kmer counting that extends 
and is more efficient than KCH. The system \ignore{implementing such an extension of 
KCH}is called \FK, and has been carefully engineered in order to 
maximize the potential of Apache Spark~\cite{Spark}, the big data framework on 
which it is 
based.  The result is that, to the best of our 
knowledge, \FK is the fastest distributed system available so far for extracting 
\kmer statistics from large genomic and meta-genomic sequences using arbitrary 
values of $k$. Indeed, attention has been payed in order to enforce a 
balanced distribution of the workload of $k$-mer statistics aggregation tasks 
over the nodes of a computing cluster. This same approach may be useful in 
other scenarios involving more general notions than \kmers, like spaced words 
and seeds (see \cite{horwege2014spaced,leimeister2014fast} and references 
therein).  The performance of \FK has been assessed though a comparative 
experimental analysis with other distributed \kmer statistics systems over 
real-world datasets. 

The rest of the manuscript is organized as follows. In Section 
\ref{sec:related_work} some background for our work is provided. In Section $3$
the algorithm \FK is described along with some implementation 
details. The results of an experimental evaluation of \FK, and an improved 
version of it, are presented in Section $5$, as well as an experimental 
comparison with other distributed 
systems for the collection of \kmer statistics. Finally, some 
conclusions and future 
directions for our work are outlined in Section \ref{sec:conclusions}.

%% file: sec_state_of_art.tex

\section{Background}
\label{sec:related_work}
Here we first summarize a brief history on the evolution of distributed 
computing, then provide some details on the \MR programming paradigm and 
finally describe the main approaches proposed for $k$-mer counting in this 
context.

\subsection{The Evolution of Distributed Computing for Big Data Management}
	
	\subsubsection{Computational Power  via Commodity Hardware}
	In the late nineties, in order  to cope with an increasing demand of 
computing  power, the design and deployment of data centers went through deep 
changes, gibing birth to the so called {\it server consolidation process}. The big 
mainframes and supercomputers left their place to small standard 19" racks with 
many x86 processors. In that period, the new data centers consisted of three 
separated components: {\it servers}, {\it storage} and {\it networking}. In 
turn, those racks could be interconnected so as to have a form of parallelism in 
terms of distributed computing, \emph{de facto} obtained via commodity 
hardware. 

Thanks to the  scalability and virtually unlimited computational  
power, those new networks offered an effective and affordable  solution to cope 
with the challenging requirements posed by the spread of the Internet.     
To give a few examples: (a) the first versions of the Google web pages index 
were  constructed via a systematic use of an approach to distributed computing 
based on commodity hardware~\cite{ghemawat2003google,dean2008mapreduce}; (b)  
Apache designed the Hadoop framework with the aim to devise a tool 
supporting the efficient processing  of large datasets, exploiting cluster 
resources through data partitioning and the \MR paradigm. Specifically,  
Hadoop was originally designed to run on many small servers with local CPU, 
memory, and storage. The availability of local resources enabled the concept of 
physical grid as a typical (at the time) Non Uniform Memory 
Access (NUMA) architecture. The communication among nodes was  provided by the Hadoop 
Distributed File Systems like a sort of a common storage where data could be 
shared (if necessary). This solution was found to be extremely flexible and 
scalable and therefore, started to be adopted by many big players. 
	
At the same time, Hadoop was installed also on sets of workstations 
(commodity hardware) providing unexpected computing power. These kind of 
clusters were able to exploit the full power of the Hadoop architecture 
providing balanced resource usage and the exploitation of data-locality\ignore{(each single task was executed inside a single 
node making the CPU closer to data stored on the local disk, and therefore 
exploiting data locality for intra-task communications)}.

Summing up, since their introduction, both \MR and Hadoop have 
become a cornerstone of big data processing. Indeed, quite remarkably, the key 
for their success is that \MR programming  supported by Hadoop provided  
programmers with a quite convenient environment to develop effective 
applications, allowing them to focus more on the specific task at hand, rather than issues such as synchronization and process-to-process communication, as opposed to what traditional, low-level primitives such that provided by MPI Standard (Message Passing Interface \cite{gropp1999}) or its ancestor, PVM (Parallel Virtual Machine \cite{geist1994pvm}), allowed for. 
Indeed, within such programming environments, concurrency must be explicitly handled by the programmer and the running program strongly depends on the 
physical network topology. In the realm of bioinformatics, this point is well 
illustrated in the Magellan Final Report~\cite{coghlan2011magellan} regarding 
the collection of $k$-mer statistics in large meta-genomic datasets: MPI 
solutions would work but it was much more convenient to use Hadoop and 
\MR, also considering the availability of higher level tools like Pig 
\cite{BioPig2015,olston2008pig}. 
	
\subsubsection{From Commodity Hardware to the Cloud: Converged 
Architectures}
The main drawbacks of the big data processing approaches outlined in the preceding section, were relative to the inherent complexity behind installing software and middleware on many nodes and manage their configuration changes.
	
With the beginning of the new millennium, the application of Virtualization 
\cite{bugnion2002system, sugerman2001virtualizing,rosenblum2005virtual} in the 
big data context promised help to speed up this process. To this end,  
manufacturers assembled pretested components, servers, storage and networking \emph{converged} in a 
single rack, in order to speed up the tuning and testing phases. Moreover, the introduction of a new software layer, namely \emph{hypervisor} with the promise of taking care of all hardware details, introduced the concept of {\it Virtual Machine} (VM). 
Hundreds of VMs became manageable with a single click, and this enabled the 
design of very large clusters with thousands of computing cores and high amounts of working memory. All these advantages brought to the \emph{converged compute and storage for virtualized environments}. 
To keep this flexibility, VMs may have access to storage areas, represented by a logical partition of a huge disk array, represented by a Logical Unit Number, or LUN, that refers, by extension, also to the partition itself. The storage is therefore deployed in Storage Area Networks (SANs), a network, separated from compute servers, which connects storage resources, typically using fiber optics cabling. Such a solution is very effective but introduces a serious scalability issue, as this connection might become a bottleneck for the entire architecture (from the cluster perspective).
	Indeed, such an approach is effective for cloud resource providers that 
offer custom made Infrastructures as a Service (IaaS), where each 
infrastructure is virtual and independent of the others. However, it introduces  
undesirable side effects when applied to distributed architectures such as 
Hadoop: although not immediately evident, the SAN represents an actual bottleneck 
in the Hadoop cluster design because in this context data locality becomes only a logical 
concept, as storage is fundamentally physically separated from compute.\ignore{That is, the corresponding physical counterpart on each node can be 
obtained using a logical partition of a single (shared) disk.} More explicitly,  
node data locality no longer exists. As a side effect, by adding nodes to the 
cluster, the I/O channel might suffer from congestion, even with a small number of physical 
clients: scalability, both in the form of response time and throughput, may be 
compromised independently of any careful design of the program being executed from the scalability standpoint. 
	To face such issues and to restore the concept of data locality as originally meant, local onboard memory caches, sometime using SSD 
technology, were specifically introduced on the CPU board. This approach mitigated 
the problem but increased the management complexity so that big players, like 
Amazon, singled out their offer of \MR  by supporting it with highly 
specialized managed services, i.e., Amazon Elastic MapReduce (Amazon EMR). In 
such a case, the provider, instead of a set of standard VMs, is offering a 
Hadoop cluster with the required resources (number of nodes, VCores, RAM and 
storage) already configured and tuned up, ready to run. In other words, the 
initial offer of a virtual infrastructure (IaaS) became a full platform 
(Platform as a Service or PaaS for short) adding the required middleware to the 
computing resources and the skill to get the best performance.
	In some cases, the services include special purpose and dedicated 
hardware resources far away from the standard architecture used for Amazon 
Elastic Web Services. Therefore such services allow customers to develop, run 
and manage their distributed applications without concerning with the complexity 
of building, tuning and maintaining the infrastructure.
	
	\subsubsection{ Storage Closer to Compute in the Cloud: Hyperconverged 
Architectures}
	
	In the past $5$ years, to leverage the concept of data locality for 
distributed clusters, manufacturers introduced the concept of Hyperconverged 
hardware. A new server generation has been designed assembling CPU, memory and 
storage (disk or SSD) in a single box, and therefore substituting the SAN and its 
disk arrays with a local distributed storage.
	This change was supported by VMWare, which added to its hypervisor new 
capabilities dedicated to Software Defined Storage (SDS), replacing the 
classic SAN with a Virtual SAN (VSAN), thus gaining a higher level of abstraction obtained by merging the storage available 
on each node, allowing for a more granular and flexible segmentation of storage resources. This change was even more effective because redundancy and 
reliability were handled at node level instead of a single cabinet for storage.
	This way data locality returned to be related to the physical 
distance between stored data and CPU that should process it, and the whole 
cluster storage returns to be a sort of NUMA architecture, allowing nodes to 
distinguish between locally or remotely stored data.
	Local data can be accessed through high speed I/O channels, while 
remote data can be accessed through slower network connection.
	Scalability is enforced because adding new boxes (more CPU, RAM and 
storage) improve the overall I/O performance providing new local I/O channels 
and without overloading the intra node network connection.
	As far as Hadoop is concerned, with this hardware architecture, data 
locality can be again exploited because each box hosts a set of cluster virtual 
nodes. All these instances can access their data which will be kept close to the 
CPU by the SDS capabilities.

	\subsection{The \MR Programming Paradigm and the Middleware Supporting 
It}

	\subsubsection{The Paradigm} \label{sec:mr}

	\MR \cite{dean2008mapreduce} is a paradigm for the processing of large 
amounts of data on a distributed computing infrastructure. Assuming the input 
data is organized as a set of \KV{key}{value} pairs, it is based on the 
definition of two functions. The {\em map} function processes an input 
\KV{key}{value} pair and returns a (possibly empty) intermediate set of 
\KV{key}{value} pairs. The {\em reduce} function merges all the intermediate 
values sharing the same \empty{key} to form a (possibly smaller) set of values. 
These functions are run, as tasks, on the nodes of a distributed computing 
cluster. All the activities related to the management of the lifecycle of 
these tasks, as well as the collection of the map results and their 
transmission to the reduce functions, are transparently handled by the 
underlying framework (\emph{implicit parallelism}), with no burden on the 
programmer side.	
	
	\subsubsection{Apache Hadoop} \label{sec:hadoop}
	
	Apache Hadoop is the most popular framework supporting the \MR 
paradigm. It allows for the execution of distributed computations thanks to the 
interplay of two architectural components: YARN (\emph{Yet Another Resource 
Negotiator})~\cite{vavilapalli2013apache} and HDFS (\emph{Hadoop Distributed 
File System})~\cite{HDFS}. YARN manages the lifecycle of a distributed 
application by keeping track of the resources available on a computing cluster, 
and allocating them for the execution of application tasks modeled after one of 
the supported computing paradigms. HDFS is a distributed and block-structured 
\filesystem designed to run on commodity hardware and able to provide fault 
tolerance through data replication. 
	
	A basic Hadoop cluster is composed  of a single \emph{master node} and 
multiple \emph{slave nodes}. The master node arbitrates the assignment of 
computational resources to applications to be run on the cluster, and  
maintains an index of all the directories and the files stored in the HDFS 
distributed file system. Moreover, it physically tracks the slave nodes, storing 
 the data blocks making up these files. Slave nodes, in turn, host a set of 
\emph{worker}s (also called \emph{containers}), in charge of running the map and 
reduce tasks of a \MR application, as well as using the local storage to 
maintain a subset of the HDFS data blocks.

	One of the main characteristics of Hadoop is its ability to exploit 
\datalocal computing. This term, refers to the possibility to move 
tasks closer to the data they need to operate on (rather than the opposite). This allows to greatly reduce network congestion and increase the overall throughput of the 
system when processing large amounts of data. Moreover, in order to reliably 
maintain files and to properly balance the load between different nodes of a 
cluster, large files are automatically split into smaller blocks, replicated and 
spread across different nodes.
	
	\subsubsection{Apache Spark}

Spark is a fast and general distributed system for cluster computing 
on big data. It provides high-level APIs which support multiple languages such 
as Scala, Java and Python. It consists of two main blocks: a programming model 
that creates a dependency graph, and an optimized runtime system which uses this 
graph to schedule work units on a cluster, and also transports code and data to 
\emph{worker} nodes where they will be processed by \emph{executor} processes.

\paragraph{Resilient Distributed Datasets}
At the core of the Spark programming model is the \emph{Resilient Distributed 
Dataset} (RDD) abstraction, a fault-tolerant, parallel data structure that can 
be created and manipulated using a rich set of operators. Programmers start by 
defining one or more RDDs through \emph{transformations} of data that originally 
resides on stable storage or other RDDs. Example of transformations are 
operations like \texttt{map}, \texttt{filter} or \texttt{reduce}, and are 
\emph{lazy}, in the sense that they return a new RDD which depends on the old 
RDD but they are not performed until an \emph{action} that involves such RDD is 
specified in the pipeline graph. Actions are operations that return a value to 
the application and export data to a storage system. 
 

Apart from the internal cluster manager, Spark applications can be run also on 
external cluster managers like Hadoop YARN \cite{vavilapalli2013apache} or Mesos 
\cite{hindman2011mesos}. Moreover, a Spark application can be run over a 
distributed file system, e.g., HDFS \cite{HDFS}. This allows each worker node 
of a cluster to read input data and to write output data using a local disk 
rather than a remote file server. 

\paragraph{Partitions, parallelism, and shuffling}
\label{parag:intro_partitions_parallelism}
By default, Spark tries to read data into an RDD from the nodes that are close 
to it. Since Spark usually accesses distributed data, to optimize transformation 
operations it creates partitions to hold the data chunks.
 The number of partitions of an RDD reflects the degree of parallelism (number 
of tasks) employed by Spark while processing it.
When an RDD is created by loading files from HDFS, its number of partitions is 
equal to the number of input splits\footnote{The size of an input split depends 
on the \emph{block size}, a configurable parameter of the \MR ecosystem.} 
of the original file on HDFS. The Spark mechanism for redistributing 
data across partitions is called \emph{shuffling}. It occurs when certain 
transformations, such as \texttt{groupByKey} or \texttt{reduceByKey}, are issued 
on an RDD and cause moving data across different processes or over the wire 
(between executors on separate machines). 
An RDD that is obtained via a shuffle transformation of another RDD, will 
inherit its number of partitions. However, as far as choosing a ``good'' number 
of partitions is of concern, what is typically desired is to have at least as 
many partitions as the number of cores. 
A way to influence this choice is by specifying a custom value for the {\tt 
spark.default.parallelism} property. Another option, is introducing a custom 
\emph{partitioner}.
Partitioners are objects that define how elements of a key-value RDD are 
partitioned by key. The Spark default partitioner (i.e., 
\texttt{HashPartitioner}) chooses the partition where to map an element as the 
Java's \texttt{Object.hashCode} value of its key (modulo number of partitions), 
or 0 for negative hashes. It is also possible to implement a custom partitioner 
class defining a custom method for assigning keys to partitions. This feature is 
useful when for some reason the default partitioner causes RDD data to be 
unevenly distributed across partitions. 

\subsection{Big Data Based Approaches for the Analysis of Biological Sequence
Datasets: The Special Case of $k$-mers Counting}
\label{sec:kmerscounting}
The modern
high-throughput technologies produce high amounts of sequence collections of 
data, and several methodologies have been proposed for their efficient storage 
and analysis \cite{Giancarlo14,Vinga14}. Recently, approaches based on \MR and 
big data technologies have been proposed (see, e.g., \cite{Cattaneo2017b}, and 
\cite{Cattaneo2017} for a complete review on this topic). An important issue in 
this context is the computation of $k$-mer statistics, that becomes challenging 
when sets of sequences at a genomic scale are involved. 
Due to the importance of this task in several applications (e.g., genome 
assembly \cite{pavel11} and alignment-free 
sequence analysis \cite{Giancarlo14,Vinga14}) many methods that use 
shared-memory multi processor architectures or distributed computing have been 
proposed. 

The basic pattern followed by most of these methods is to maintain a 
shared data structure (typically, a hash table) to be updated according to the 
\kmers extracted from a collection of input files by one or more concurrent 
tasks. When memory is not enough to maintain all the extracted \kmers, these 
can 
be organized in disjoint partitions and temporarily saved on file without 
aggregation. Then, they will be loaded in memory one partition at time and 
summed to return the definitive \kmer statistics.

Here, we provide a 
summary of the main techniques proposed for $k$-mers counting in the biological 
scenario, distinguished in two main categories: those designed to work on 
shared memory and/or multi-processor systems, and those implemented for 
distributed systems (the interested reader can find a more deep survey of them 
in \cite{KCH}).

\paragraph{Shared Memory, Multi-Processor Systems}
MSPKmerCounter~\cite{li2014mspkmercounter} introduced a disk-based approach 
where consecutive \kmers are not saved individually but first compressed to a 
single \emph{superkmer}. 
This solution leads to a significant reduction in the amount of data to be temporarily saved on disk and, then, recovered to memory, thus allowing for a significant performance boost with respect to other algorithms. 
The notion of minimizer has been refined in  KMC2~\cite{deorowicz2015kmc} and, 
later, in  KMC3~\cite{doi:10.1093/bioinformatics/btx304} with the introduction 
of $k$-mer \emph{signatures}. These are a specialization of minimizers and are 
built with the idea of discouraging an extremely imbalanced partitioning of 
super $k$-mers among the different bins while keeping the overall bins size as 
small as possible. An additional contribution provided by these systems is in 
the counting phase. Input super $k$-mers are broken into $(k, x)$-mers, a 
compact representation of a sequence of \kmers, and sorted efficiently using a 
parallel version of radix sort~\cite{Cormen2001}. 

\paragraph{Distributed Systems}

The applicability and scalability of multi-processor shared-memory architectures 
is inherently constrained by architectural factors, such as the maximum number 
of processing cores on a processor, and the maximum amount of memory on a 
single hosting machine. Distributed systems allow to overcome these limitations. 
Indeed, the availability of an arbitrary number of independent computation nodes 
allows to virtually extend to any size the data structure used to keep the 
$k$-mer statistics in memory, while using the network as a temporary buffer 
between the extraction phase and the aggregation phase. 

This is the approach followed by Kmernator~\cite{kmernator2012} and 
Kmerind~\cite{Kmerind2016}. 
Both these tools are developed as MPI-based parallel applications and are able 
to handle data sets whose size is proportional to the overall memory of the 
MPI-based system where they are run. However, the development and management 
of an in-house MPI-based supercomputing facility is usually very complex and 
expensive. 

An alternative approach that is gaining popularity in the 
community of Bioinformaticians is the usage of big data processing 
frameworks, such as Apache Hadoop and Spark. As already mentioned, these 
technologies are cheaper and adopt a simpler 
programming model than MPI. To cite only some examples, 
BioPig~\cite{nordberg2013biopig} is an Hadoop-based analytic toolkit for the 
processing of genomic data. It has been developed as an extension of the Pig 
language that, in turn, offers a set of data querying and transformation 
primitives that are translated into \MR jobs. BioPig includes a module, called 
pigKmer, that allows to extract and count the \kmers existing in a set of 
sequences. Each sequence is splitted in several blocks saved on the different 
nodes of a distributed system, with each block being processed by a distinct 
task. The \kmers extracted this way are then aggregated, using a reduce 
operation, and finally counted. An alternative distributed \kmers counter is 
the one provided by ADAM~\cite{ADAM}, a Spark-based toolkit for exploring 
genomic data, which follows the same application pattern of BioPig. The 
algorithmic approach of these two systems is pretty naive, so they are able to 
process very large genomic sequences but at the expense of very bad 
performance.

The first and, to date, the only distributed system able to extract efficiently 
\kmer statistics from large collections of genomic sequences, with $k\leq31$ is 
KCH\cite{KCH}. It is a distributed system based on \MR and Hadoop 
which follows a two-level aggregation strategy. In particular, it 
first partitions the universe of possible \kmers in a fixed number of bins 
($291$, by default), and then extracts the \kmers counts from a collection of 
input sequences in two stages of alternate map and reduce 
tasks. In the first stage, each map task creates a distinct hash table for each 
bin and updates them with the statistics of the \kmers extracted from a chunk 
of the input sequences. At the end of this stage, each map task returns its 
collection of hash tables holding the partial \kmer counts. During the second 
stage, all the hash tables corresponding to the same bin are aggregated by a 
reduce task and the result is saved on file.  This strategy is able to 
significantly reduce the communication overhead between the different nodes of 
the system, thus allowing for execution times that are up to $100\times$ faster 
than those of BioPig, when run on fairly large sequences.

%% file: sec_approach.tex

\section{Methods}

\subsection{Basics}
\label{sec:background}

\noindent Let  $\Sigma$ be an alphabet and $S$ be a finite set of collections 
of sequences over $\Sigma$. A  {\it cumulative 
statistics} collects how many times each of the $k$-mers in $\Sigma^k$ appears 
in the collections  of sequences  in $S$.  Here $\Sigma = \{A, C, G, T \}$  and 
$S$ is a collection of genomes or meta-genomes. 

The algorithms for \kmer statistics computing usually have a common 
structure: they first process the sequences in $S$  from left to right in order 
to extract all \kmers, and then perform aggregation and evaluation. A naive 
implementation, such that all single \kmers are extracted in a sliding window 
fashion, is highly redundant in space. Indeed, for an input length of $n$ 
characters, generating all \kmers determines an unfolded sequence of 
$(n-k+1)\cdot k$ symbols. Since, by definition, consecutive \kmers along a 
sequence share $k-1$ symbols, it would be beneficial to have a compressed 
representation of them, where all contiguous \kmers are stored in a compact 
sequence. Yet unfortunately, to be able collect the statistics, especially in a 
distributed setting where different portions of the input data will be processed 
in physically separated machines, we need a way to keep together all instances 
of each unique \kmer for the evaluation phase. A clever solution to this problem 
is based on the notion of 
minimizers~\cite{roberts2004reducing,li2014mspkmercounter,Deorowicz15}.

\paragraph{Minimizers}

Given a \kmer $s$, a {\it minimizer} of $s$ is a word  of length $m$ (with $m$ 
fixed a priori) occurring in $s$. Usually many consecutive 
\kmers 
have the same minimizer and, therefore, can be compressed into a sequence of 
more than $k$ symbols, a \emph{superkmer}, significantly reducing the 
redundancy. 

Minimizers may be used to partition \kmers into multiple disjoint 
sets, as well as retaining adjacent \kmers in the same partition: superkmers can 
be distributed into different bins according to their related minimizer ensuring 
that all the corresponding instances of a \kmer will appear in the same bin (see 
Section \ref{subsec:extracting_superkmers} for further details).

%

\subsection{$k$-mer Statistics Collection on Spark: the \FK Algorithm}
Here \FK is described, focusing also on the engineering aspects which make it 
more efficient with respect to its competitors (e.g., KCH).

\subsection{Design Overview}

\begin{figure}[t]
	\centering
	\includegraphics[width=0.9\textwidth]{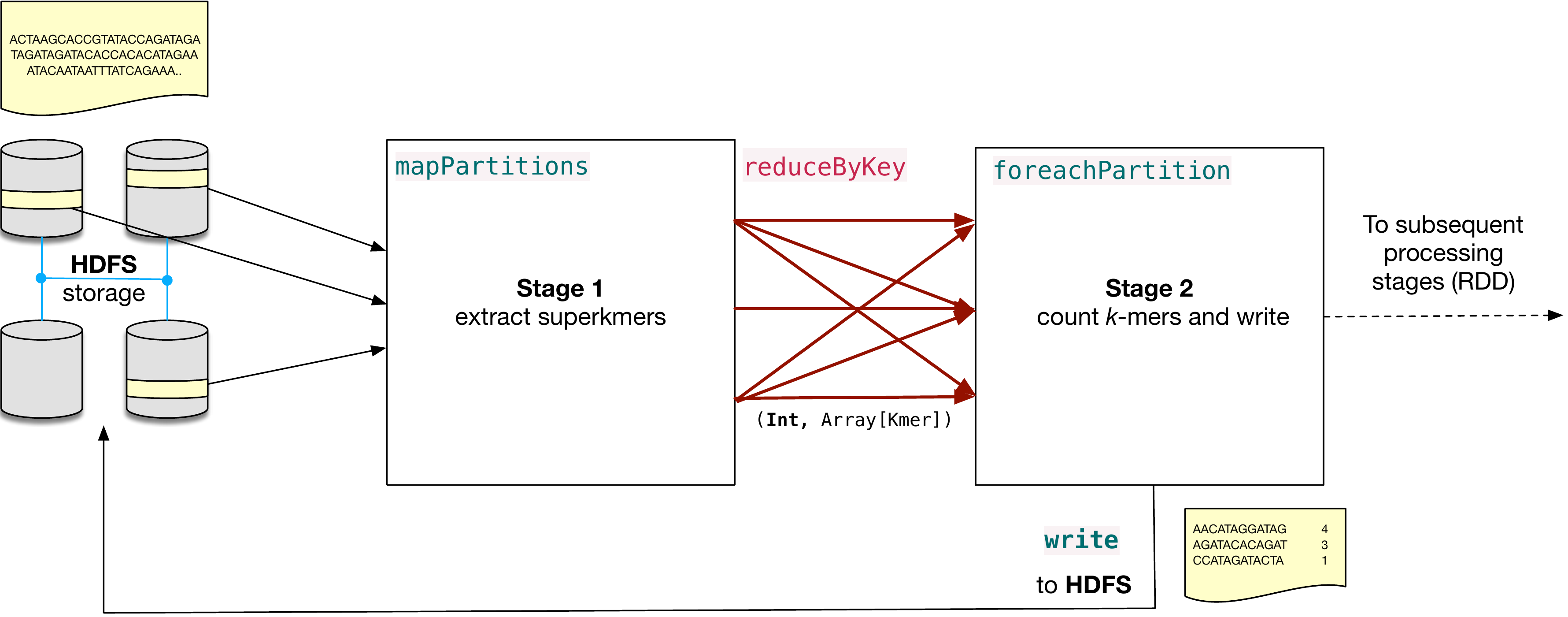}
	\caption{Stages of the pipeline implementing \FK.}
	\label{fig:pipeline}
\end{figure}

The \FK algorithm is implemented on the Spark pipeline described in
Figure~\ref{fig:pipeline}. The figure does not show the initial 
configuration/parameterization stage. The core of the pipeline consists of two 
main stages, as well as a preliminary stage responsible of fetching the input 
dataset, that is a FASTA/FASTQ file \cite{fasta2016, cock2010sanger}, from HDFS 
storage, and delivering all of its blocks to the first stage of the pipeline 
(leftmost portion of Figure~\ref{fig:pipeline}). The first stage performs the 
extraction of superkmers. The second stage computes and collects the 
\kmer statistics. Both stages are described in detail in the following.

\subsubsection{First stage: extracting superkmers}
\label{subsec:extracting_superkmers}
\begin{figure}[ht]
\centering
		\includegraphics[width=.60\textwidth]{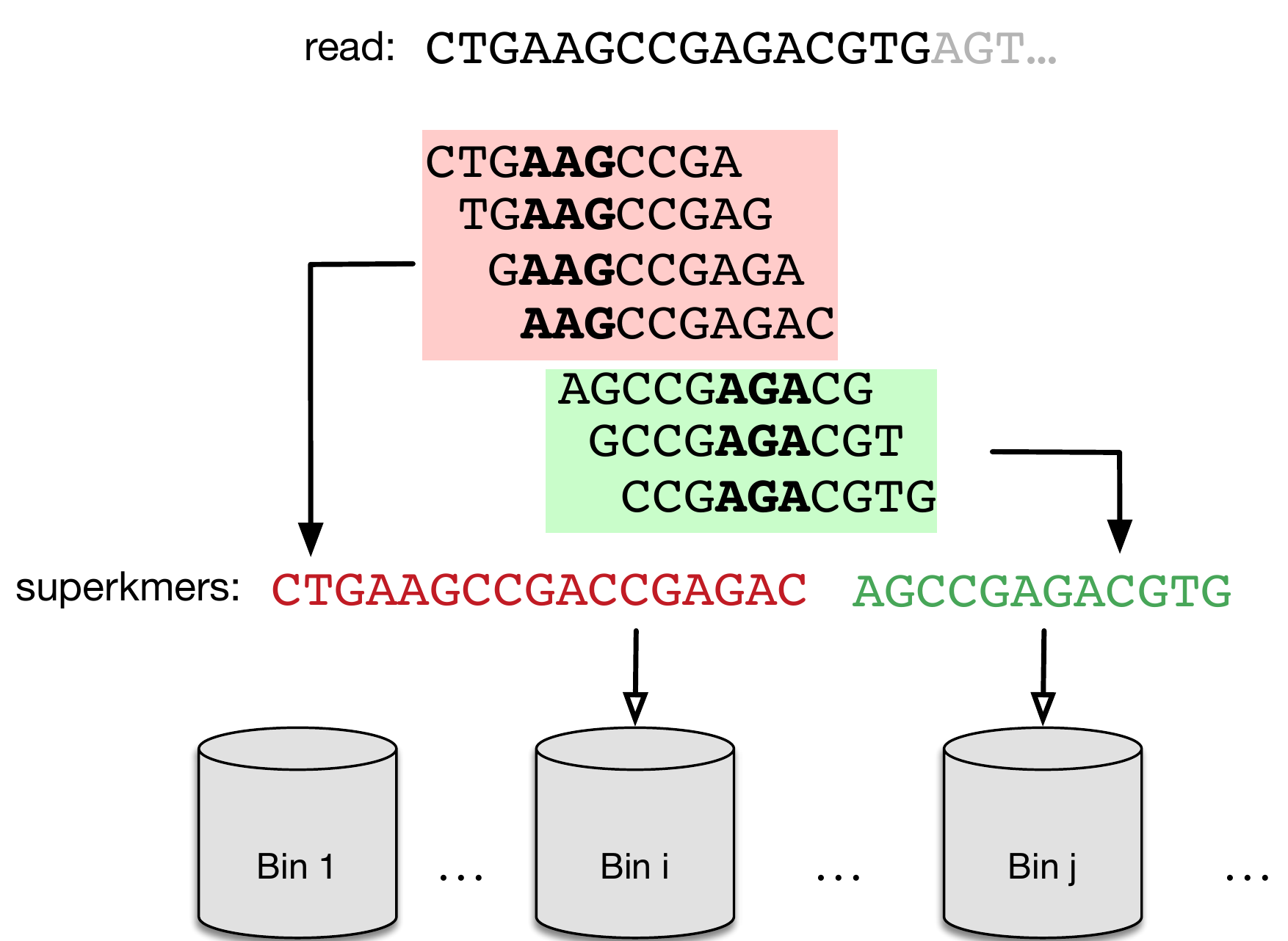}
			\caption{Extraction of superkmers from input sequences, using signatures\\ ($k=10,m=3$).}
			\label{fig:superkmers}
		\end{figure}

Since consecutive \kmers share $k-1$ symbols, an approach that 
exhaustively extracts all \kmers with a sliding window and propagates them to 
the next stage leads to high redundancy. To address this issue, the first stage 
of our approach processes all input sequences in a way to guarantee a degree of 
compression: sequences are broken into superkmers using their corresponding 
signatures which are in turn used to implement a binning mechanism. This is a 
variant of the Minimum Substring Partitioning (MSP) 
technique~\cite{li2013memory,li2015mspkmercounter}, which is in turn based on 
the notion of minimizers. In particular, \FK adopts a slightly different notion 
that is the one of \emph{signatures}~\cite{deorowicz2015kmc}, \ie canonical 
minimizers of length $m$ (a tunable parameter) that do not start with \emph{AAA} 
nor \emph{ACA}, neither contain \emph{AA} anywhere except at their beginning. 
A toy example of splitting a sequence into superkmers using signatures is depicted in Figure~\ref{fig:superkmers}.
\paragraph{From signatures to bins} Since signatures are based on  lexicographic 
ordering, superkmers having a given signature $s$ are mapped to one of a set of 
$B$ bins (a parameter) using a shift-based integer hash function, thus aiming at 
a uniform distribution of superkmers in processing units for the subsequent 
phase.

The output of the first stage is therefore a sequence of bins,  where each bin 
is described by an integer key in the range $\{1,\dots,B\}$ and holds a sequence 
of superkmers. Then, bins originating from different distributed workers are 
automatically aggregated by Spark based on their key in an intermediate phase 
before the next stage (red shuffling phase in Figure~\ref{fig:pipeline}).

\subsubsection{Second stage: partitioned counting}
\label{subsec:partitioned_counting}
The second stage is responsible of the counting phase: due to the 
signature-based binning process, all instances of a given \kmer are guaranteed 
to reside in the same bin. Therefore each bin is processed independently and all 
the \kmers contained therein are inserted in a hash table that also maintains 
their relative counts. After processing each bin, the table is traversed, and 
counts are stored on HDFS.  


\subsubsection{Implementation details}


\FK has been implemented as a Spark application using 
Scala~\cite{odersky2004scala}.
%
Input sequences are read by \FK using FASTdoop~\cite{ferraro2017fastdoop}.
%
%
%
\FK represents \kmers (and likewise, superkmers) of arbitrary length using  an 
\texttt{Array[Long]}, where information is encoded by a binary 
representation. Therefore, since the alphabet of valid nucleotides consists of 
four items, as $\Sigma=\{A, C, G, T\}$, a string over $\Sigma^k$ can be encoded 
using only two bits per symbol.

\section{Results and Discussion}
\label{sec:exp1}

Here we describe the results of an experimental analysis which show as 
different choices of the parameters and Spark-related configurations may
result in different performance of the algorithm presented as a case study. 

\subsection{Setup}

%
%

\paragraph{Testing Platform}
The experiments have been performed on the Microsoft Azure Cloud infrastructure.
In particular, a $8$-node Spark $2.1.0$ cluster has been deployed into 
HDInsight, Microsoft's cloud distribution of the Hadoop ecosystem (Hadoop 
$2.7.3$), based on the Hortonworks Data Platform (HDP). Two cluster nodes act as 
head nodes, and are equipped with an $8$-core $2.4$ GHz Intel Xeon E$5$-$2673$ 
v$3$ processor and $28$GB of RAM each, plus other six worker nodes, each with 
two $8$-core $4.78$ GHz Intel Xeon E$5$-$2673$ v$3$ processors for a total of 
$16$ cores, $112$GB of RAM and a $800$GB local SSD disk, and an overall disk 
capacity of $4.8$TB. 
The job configuration consists of $16$ Spark executors 
with $2$ cores each (for a total of $32$ workers). 


\paragraph{Datasets}
The dataset considered here refers to~\cite{hess2011metagenomic}. In 
particular, the $SRR094926$ run of the $SRP004875$ SRA study 
(available on the NCBI short read archive) has been used, for a total 
occupation of about $125$GB (FASTA format). For tuning experiments, a different 
file containing only the first $32$GB of this run has been considered.  See 
Table~\ref{tab:dataset_sizes} for summary information.

\begin{table}[ht]
\centering
\begin{tabular}{@{}rrrrr@{}}
\cmidrule(l){2-5}
\multicolumn{1}{l}{} & \multicolumn{2}{c}{\textbf{$k=28$}} & \multicolumn{2}{c}{\textbf{$k=55$}} \\ \cmidrule(l){2-5} 
\textbf{kmers} & \multicolumn{1}{c}{\textbf{32GB}} & \multicolumn{1}{c}{\textbf{125GB}} & \multicolumn{1}{c}{\textbf{32GB}} & \multicolumn{1}{c}{\textbf{125GB}} \\ \midrule
\textbf{distinct} & 12,551,234 K & 37,337,258 K & 14,203,028 K & 47,830,662 K \\
\textbf{total} & 22,173,612 K & 86,674,803 K & 18,722,642 K & 73,209,044 K\\
\end{tabular}
\caption{Number of distinct and total $k$-mers for our datasets.}
\label{tab:dataset_sizes}
\end{table}

\subsection{Experimental Evaluation}
Here a study of different configurations and parameters, as well 
as of their main implications on the performances of \FK, is presented.

\paragraph{Value of $k$} For all experiments, we examine the running time 
performance of the \kmer statistics collection task on our datasets for two 
different, reference values of $k$: $28$ and $55$. 
\paragraph{Signature length}
Preliminary experiments have been performed in order to tune the signature 
length parameter $m$ (data not shown but available upon request). The result is 
that small values of $m$ increase the probability that consecutive \kmers 
share the same minimizer and thus reduces the I/O cost at the end of the first 
stage. However, if too small, it might render the distribution of partition 
sizes skewed and the largest partition might not fit in memory. On the other 
hand, a large $m$ will make the distribution of partition sizes evener at the 
cost of a higher redundancy (with no compression for $m \rightarrow k$). 

The assignment which yields, on average, the best 
performance on the considered datasets is $m=10$. This is in line with the 
results in \cite{deorowicz2015kmc,li2013memory} 
on datasets of comparable characteristics. 
\paragraph{Number of bins}
The number of bins $B$ used for the signatures binning scheme is considered. At 
the starting of the second stage, each partition contains a number of bins to 
be processed. Having few bins decreases the overall memory management overhead 
to be paid at the beginning and at the end of the processing of each bin, yet 
having few, very large bins which might exceed the memory available to a worker 
process. On the other hand, a larger number of bins allows for a better 
granularity of the distributed execution, as each bin can be processed 
independently, while incurring in an increased memory management overhead.

\paragraph{Spark parallelism} Aside from the parameters of the proposed 
algorithm, there is a Spark-specific parameter which may have an impact on the
running time and cluster usage, that is Spark \emph{parallelism level} 
($p$). This parameter corresponds to the number of tasks that are spawned by 
Spark (as well as the number of partitions, see 
Section~\ref{parag:intro_partitions_parallelism}), and has a side effect on the 
number of bins mapped to partitions: if bin numbers are uniformly spread, each 
task will receive a number of bins that tends to $B/p$. 

\paragraph{On large bins}
As previously stated, when using the minimizer-based approach, the distribution 
of superkmers associated to signatures can be very uneven, with small (in terms 
of lexicographic ordering), or particularly frequent signatures tending to have 
a very large fraction of superkmers, compared to the other ones. This is partly 
mitigated by the choice of signatures within a suitably filtered sets of 
canonical minimizers that do not start with common prefixes (see 
Section~\ref{subsec:extracting_superkmers}), and by the hash-based mapping of 
signatures to bins. Nevertheless, since the scheme is data-oblivious, it might 
still produce large bins. In our distributed setting, this is particularly 
relevant because they can introduce bottlenecks where the running time is 
impacted by a few number of workers that take more time than the others, thus 
leading to a non-optimal utilization of the cluster.

\begin{figure}[ht]
		\centering
		\includegraphics[width=.95\textwidth]{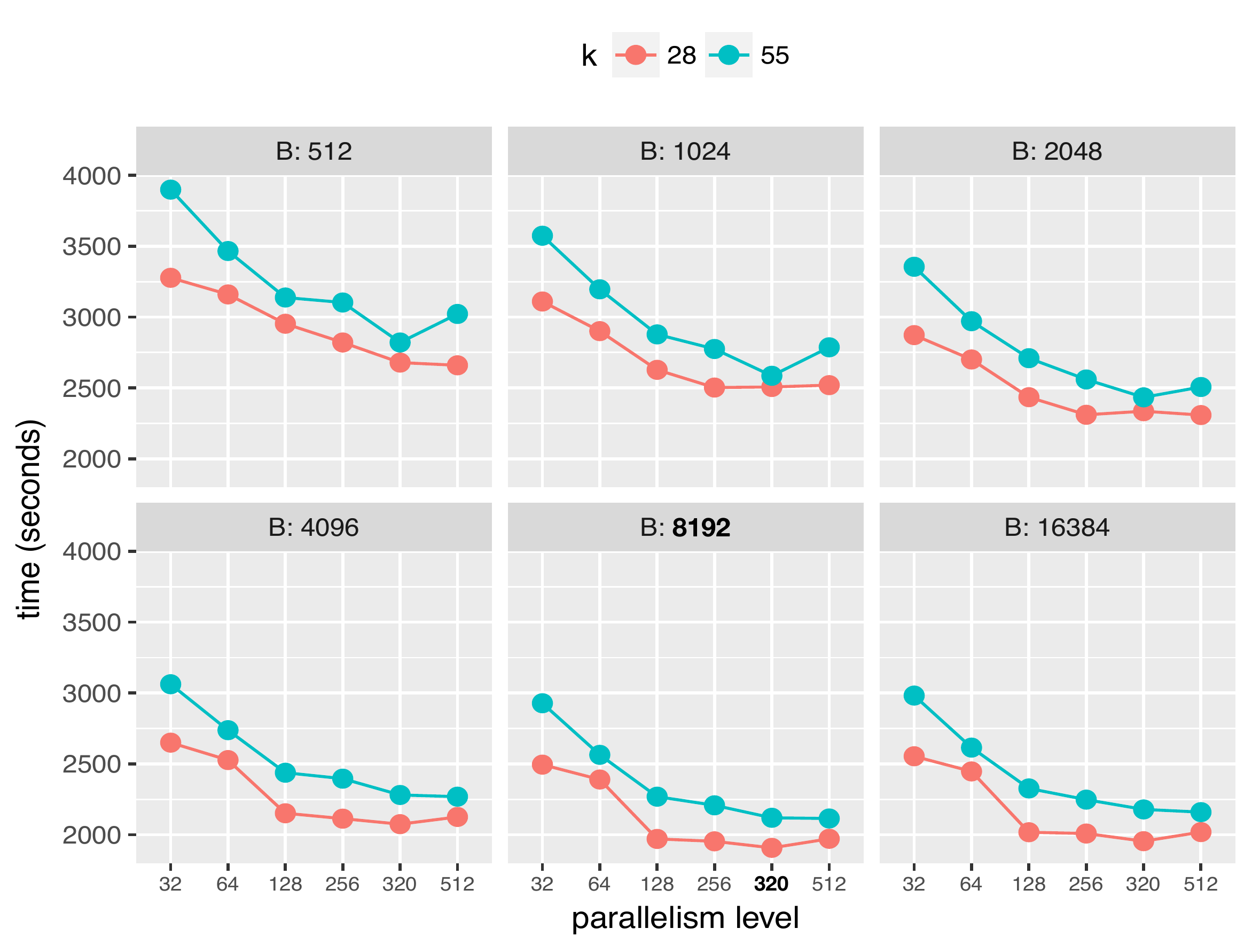}
			\caption{Algorithm execution times on the {\em 32GB} dataset with $k \in \{28,55\}$ and an increasing number of bins ($B$) and parallelism level. Best combination shown in bold.}
			\label{fig:bins_tuning}
		\end{figure}

Figure \ref{fig:bins_tuning} shows the running time of  \FK for 
values of $B$ corresponding to powers of two between $512$ and $16.384$ and a 
parallelism levels ranging from $32$ to $512$ (corresponding, respectively, to 
$1$ and $16$ average tasks per core). As for the number of bins, 
performances improve consistently for values of $B$ up to $8192$, after which we 
have no improvement (also for higher values of $B$, not plotted for legibility). 
The x-axis shows the parallelism $p$: again, for both values of $k$, we see a 
performance increase when we raise $p$ up to $320$ tasks. No improvement, if not 
a slight deterioration, is noticeable for higher values of $p$. This is expected 
as, whilst higher parallelism tends to better spread big bins across many 
partitions, conversely more tasks determine more scheduling overhead for their 
start and finalization. Spark tuning 
guidelines\footnote{https://spark.apache.org/docs/latest/tuning.html} suggest 
having a number of tasks that is in the range of $2-3$ tasks per core: this 
choice of $p$ leads to $10$ tasks per core in our test instance ($32$ workers), 
suggesting a possible heavy scheduling-related overhead.

Intuitively, increasing both values of $B$ and $p$ mitigates the big bins 
problem, as: {\it (i)} mapping signatures to more bins means potentially better 
spreads big signatures, {\it (ii)} the number of tasks better spreads big bins 
as they are more granular and distributed across more tasks, yet it does not 
fully remedy the fact that bins can have very different sizes.

A further inspection of the distribution of the single task running times for 
low values of $B$ and $p$ showed that some tasks take much longer than others 
(peaking at as much as $50\%$ of cluster underutilization, for some 
configurations). With larger values of $B$ and $p$, and in particular for the 
best configuration $B=8192$ and $p=320$, the problem is indeed mitigated, yet 
still we have a single task running at the end of the job for about $5\%$ of the 
running time.

Another problem is related to the fact that the optimal values of $B$ and $p$ 
ultimately depend on the dataset. We wish to have a solution that allows for a 
degree of adaptability of the algorithm to dataset variability, and, more 
generally, exhibiting better load balancing guarantees. In the next section, we 
explore a promising direction of improvement to address these issues.



%

	\subsection{Comparative Experimental Analysis}

\label{sec:exp2}
Here an experimental comparison of \FK  against other big 
data systems for the extraction of \kmer statistics is presented. For each 
system described in Section 
\ref{sec:kmerscounting}, the runtime configuration suggested by their 
respective authors has been adapted to the considered testing platforms.

\subsubsection{Testing Platform}

The testing platform used for this analysis is a $33$ nodes Linux-based Hadoop 
cluster, with one node acting as \textit{resource manager} and the remaining 
nodes being used as workers. Each node of this cluster is equipped with two 
$8$-core Intel Xeon $E3-12@2.40$ GHz processor and $64$GB of RAM. Moreover, 
each node has a $200$ GB virtual disk reserved to HDFS, for an overall capacity 
of about $6$ TB. All the experiments have been run using the Hadoop $2.8.1$ and 
the Spark $2.2$ software distributions.

\subsubsection{Running Times}
\label{subsec:times}

In this test we considered both the $125$GB and $32$GB datasets while using 
$k=28$ and $k=55$. As presented in Table~\ref{table:comparative}, we were not 
able to run ADAM on our testing platform because of memory issues. A further 
investigation revealed that this system extracts \kmers from an input FASTA file 
by first converting it in another format through an operation that requires the 
whole file to be loaded in the main memory of the driver process. This approach, 
apart from being extremely inefficient, prevents the system to work whether the 
driver process has not enough memory to fulfill this operation (like in our 
case). 
As expected,  the performances of BioPig are considerable lower than those of 
\FK and KCH, taking more than $10$ hours to complete, on the $125$GB dataset. 
Indeed, the lack of any aggregation strategy during the \kmers extraction phase 
and the choice of a standard character-based encoding for the extracted \kmers 
increases significantly the amount of data to be moved from the extraction phase 
to the evaluation phase, thus putting a heavy burden on the overall execution 
time of this system. 

We now turn to the case of KCH. We recall that this system has been developed to 
only support values of $k$ smaller than $32$. As for the case of $k=28$, we 
notice that \FK is about $20\%$ faster than KCH when processing the $32$GB 
dataset. This difference becomes even more significant when considering the 
$125$GB dataset. Here, \FK is about two times faster than KCH. To explain this, 
consider that KCH aggregates \kmers in bins at a much coarser level and that it 
lacks a scheduling strategy able to ensure an even distribution of the workload 
among the nodes of the underlying distributed system. 

\begin{table}[ht]
\centering
\begin{tabular}{@{}rrrrr@{}}
\cmidrule(l){2-5}
\multicolumn{1}{l}{} & \multicolumn{2}{c}{\textbf{$k=28$}} & \multicolumn{2}{c}{\textbf{$k=55$}} \\ \cmidrule(l){2-5} 
\textbf{Algorithm} & \multicolumn{1}{c}{\textbf{32GB}} & \multicolumn{1}{c}{\textbf{125GB}} & \multicolumn{1}{c}{\textbf{32GB}} & \multicolumn{1}{c}{\textbf{125GB}} \\ \midrule
\textbf{FK} & 23 & 82 & 38 & 119 \\
\textbf{KCH} & 28 & 196 & \textbf{--} & \textbf{--}\\
\textbf{BioPig} & 122 & Out of time & 450 & Out of time\\
\textbf{ADAM} & ~Out of mem & ~Out of mem & ~Out of mem & ~Out of mem
\end{tabular}\caption{Running time (minutes) for various distributed \kmer 
counting algorithms, with a time limit of $10$ hours. Dash symbols represent 
combinations where the value of $k$ is not supported by the algorithm.}
\label{table:comparative}
\end{table}

\subsubsection{Scalability}

\label{subsec:scalability} 
A comparison between \FK and of KCH is presented here, with respect to cluster 
scale and dataset sizes.
\begin{figure}[ht]
		\centering
		\includegraphics[width=.85\textwidth]{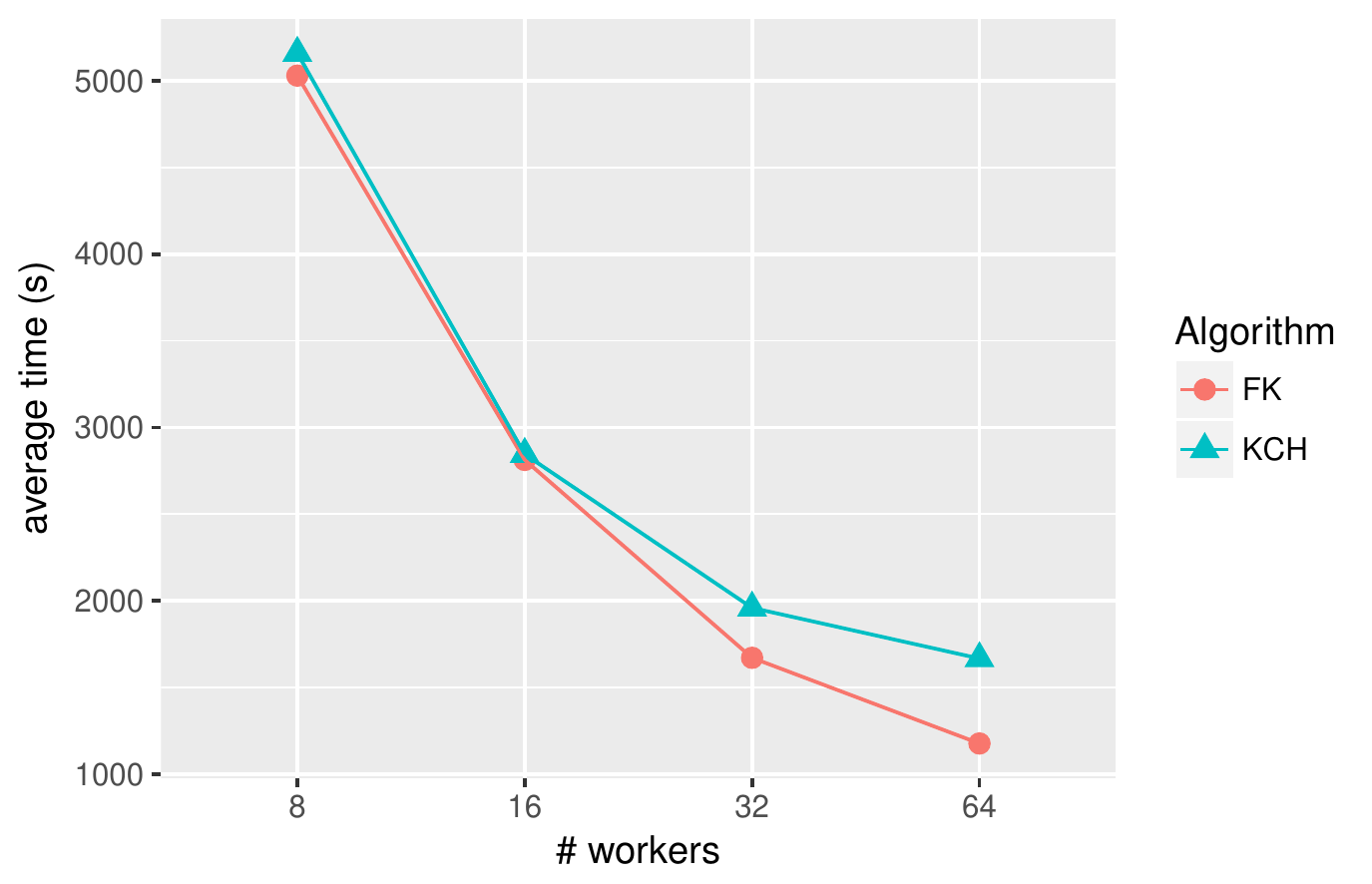}
			\caption{Running time comparison of FK and other algorithms on the {\em 32GB} dataset with $k$ fixed to $28$ and an increasing number of executors.}
			\label{fig:scalability_K28_32GB}
		\end{figure}
\begin{figure}[ht]		
		\centering
		\includegraphics[width=.85\textwidth]{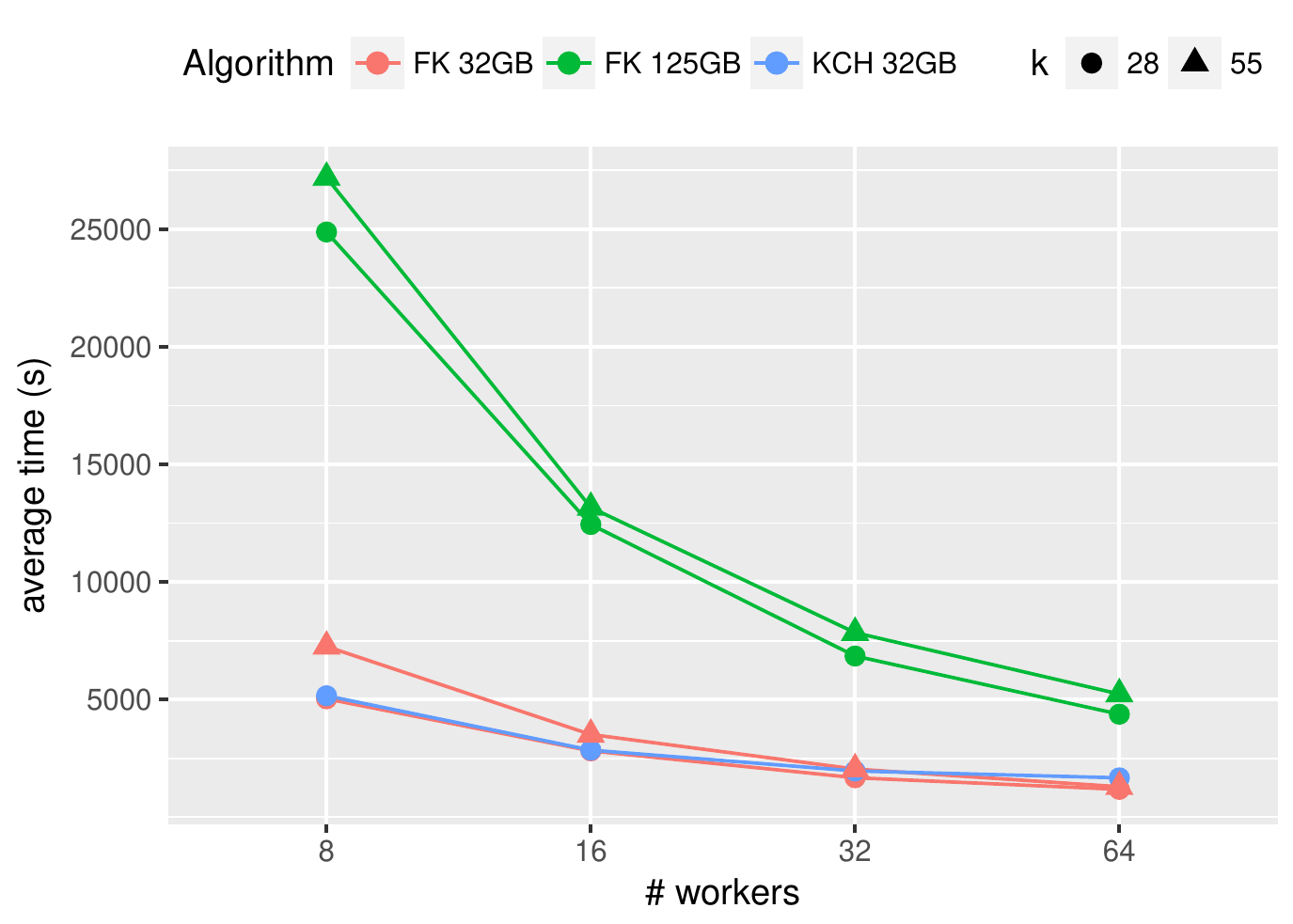}
			\caption{Running time comparison of FK and other algorithms for various values of $k$ and an increasing number of workers.}
			\label{fig:scalability_ALL}
		\end{figure}
Figure ~\ref{fig:scalability_K28_32GB} shows the running time 
comparison of \FK and KCH for $k=28$ (again, $k=55$ is not considered as KCH 
supports only values up to $k=31$). From Figure~\ref{fig:scalability_ALL} we see 
that \FK outperforms previous distributed approaches in terms of running time,  
showing to scale better for an increasing number of workers.

%% file: sec_refined_approach.tex

\subsection{Coping with Data Skew: a Multiprocessor Scheduling Inspired Partitioner}


\label{subsec:data_skew}

As previously stated, the unbalanced partitioning of bins resulting from our 
experiments is mainly driven by the minimizer-based scheme. Bins exacerbate this 
fact even more, as they contain multiple signatures, (and possibly many of such 
``big'' ones), with the consequence that few larger bins lead some workers to 
have a much longer running time than the others.
The standard partitioner of Spark does not come  into rescue, as it maps bin ids 
to partitions following their \texttt{hashCode} 
(Section~\ref{parag:intro_partitions_parallelism}), and therefore cannot take 
into consideration their size.

The necessity of achieving a balanced distribution of the workload induced 
by bins while taking into account the actual number of available processing 
units can be framed as an instance of the more general  \emph{Multi-Processor 
Scheduling (MPS)} problem ~\cite{johnson1985np}. In this problem, the input 
consists of $m$ identical machines and $n$ jobs, where each job has a processing 
time greater than zero. The goal is to assign jobs to machines so as to minimize 
the maximum load of a machine (\ie the sum of the processing time of its jobs) 
which, as all the machines operate in parallel, can be seen as the actual 
schedule time. Computationally, MPS is NP-Hard, therefore \FK resorts to an 
heuristic that is often used and simple to implement,  the \emph{Longest 
Processing Time} (LPT) algorithm. LPT proceeds by first sorting jobs by 
processing time and then assigns each of them to the machine with the earliest 
end time (i.e., lowest load) so far. This algorithm achieves an upper bound of 
$\left( \frac{4}{3} - \frac{1}{3m} \right) OPT$~\cite{graham1969bounds}.

In this setting, jobs correspond to bins, and their processing time is 
estimated as the number of \kmers contained in the corresponding bin (later, bin 
\emph{size}), and the number of machines is assumed to be the number of 
partitions.  From the \FK viewpoint, the integration of this scheduling 
algorithm requires two main modifications to the original pipeline. At the 
beginning of the computation, a new preliminary stage is run to derive an 
estimation of bin sizes by examining a sample of the input data (whose size can 
be specified, defaulting to $1\%$). This estimation is then used to compute a 
schedule using LPT. In turn, the resulting schedule is used by a custom 
partitioner replacing the original one available with Spark for mapping bins to 
partitions at the end of the first stage.


\paragraph{Granularity of working units}
\label{subsubsec:granularity}

In order to further mitigate the ``big bins problem'', we also take into 
consideration a variant allowing only one signature per bin, i.e., using the 
signature value to implement the binning process. In our multiprocessor 
scheduling analogy, the set of superkmers belonging to a signature represents a 
job. 
This choice achieves two major benefits: {\it (i)} it allows to get rid of a 
parameter ($B$), {\it (ii)} it allows for the finest granularity of work units 
for the second phase of the task, that will prove to be particularly convenient 
for our custom-partitioner based implementation, as shown in the following 
experimental section.

\noindent The next section reports an analysis of the performance improvement of our algorithm with custom partitioning.

\subsubsection{Results}\label{subsec:refined_experimental_analysis}


%

\begin{figure}[ht]
		\centering
		\includegraphics[width=.95\textwidth]{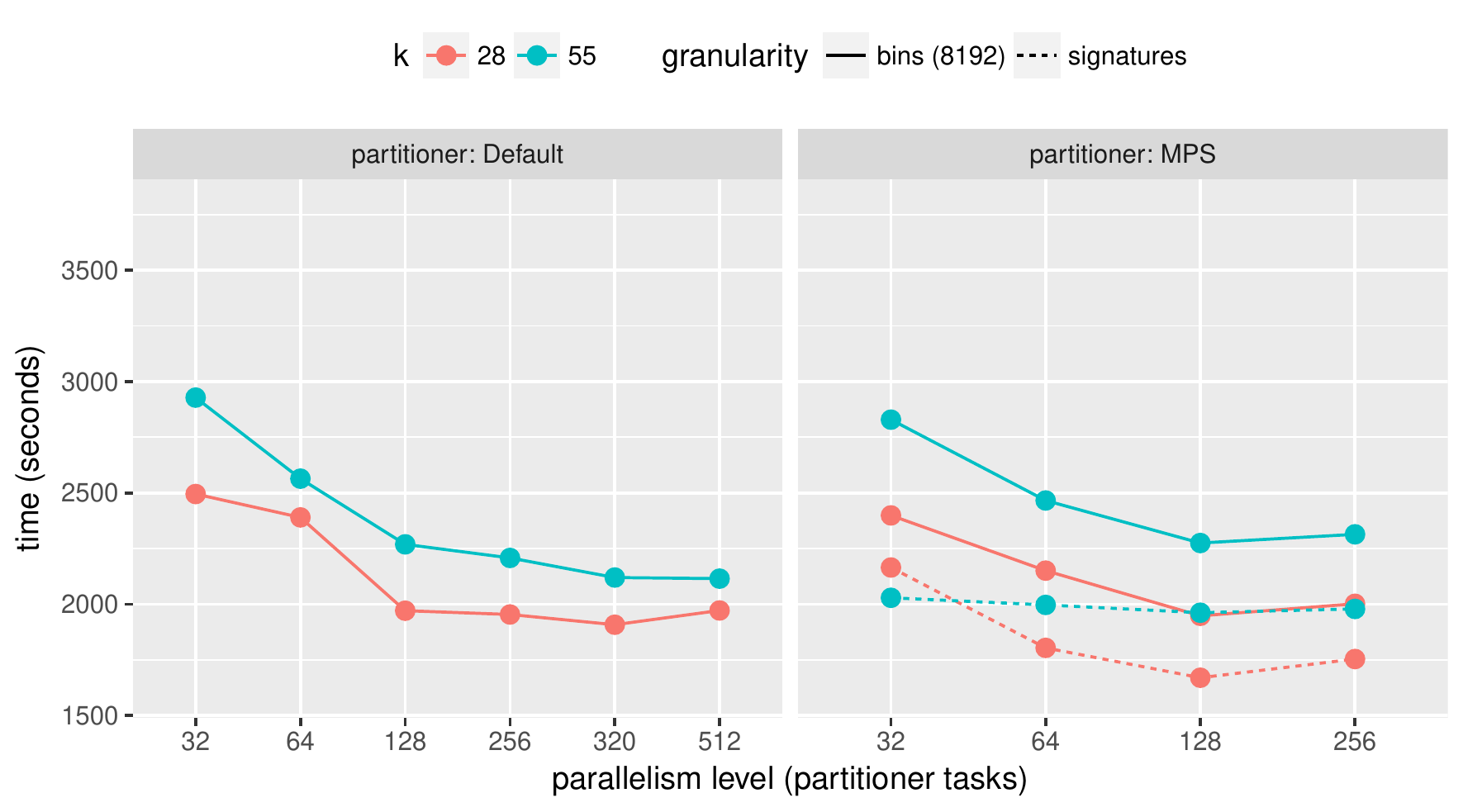}
			\caption{Comparison of execution times of \FK on the {\em 32GB} for $k \in \{28,55\}$, using default (left) or custom (right) MPS-based partitioning method, for an increasing parallelism level. The number of bins is set to 8192. For the custom partitioning scheme the performance of the signature granularity (as proposed in Section~\ref{subsubsec:granularity}) is also shown, marked with a dashed line.}
			\label{fig:partitioner_comparison}
		\end{figure}

Figure~\ref{fig:partitioner_comparison} reports a running time comparison 
between an implementation of \FK using the default Spark partitioner (left), and 
another one implementing our custom multiprocessor scheduling-based partitioning 
scheme (right) run on our $32GB$ dataset.

As for our custom partitioner,  it further compares two different granularities 
for the work units: \emph{bins} and \emph{signatures} (solid vs dashed lines). 
For the bins granularity ($B=8192$, resulting from the previous analysis), the 
impact of the custom partitioner is moderate. For the signatures choice, 
instead, the improvement of the custom partitioner has a consistently higher 
margin, suggesting an important impact of the imbalanced signatures 
distribution. On a related note, it can be noticed that the improvement is 
starting at the lowest level of parallelism ($1$ task per CPU core), and 
increases up to $128$ total tasks ($4$ tasks per CPU core).
After $128$ tasks we see no improvement: this is also expected as the goodness 
of a LPT schedule decreases with higher number of machines with respect to the 
optimal solution (in accordance to LPT bounds with respect to the optimum 
solution, see Section~\ref{subsec:data_skew}).
  

Based on these results, our default implementation makes use of our custom partitioning scheme, with signature-based binning.

%% file: sec_conclusions.tex
\section{Conclusions and Future Directions}

\label{sec:conclusions}
It is worth to remark that the advantages of technologies like Hadoop or Spark 
for the analysis of big data come at a cost. A naive usage of these 
technologies 
may bring to solutions that, although being able to run on big data, are 
inefficient. 

In this paper,  we have presented \FK, an efficient system for the extraction of 
\kmer statistics from large collection of genomic and meta-genomic sequences 
using arbitrary values of $k$. \FK succeeds in being, to the best of our 
knowledge, the fastest \kmer statistic distributed system to date, not only 
because it implements a clever algorithm for the extraction and the aggregation 
of \kmers, but even because it has been purposely engineered and tuned so to 
extract the most from the underlying Spark framework. This is especially the 
case of the different strategies that we developed for the distribution of the 
\kmers aggregation workload over a Spark cluster and that can be used as well in 
more general Bioinformatics application scenarios. 

As a future direction, we observe that the internal architecture of \FK has been conceived so as to make it easy to integrate its workflow in more complex data processing pipelines. For instances, we cite the case of distributed alignment-free algorithms. These could use \FK as a sub-pipeline to extract the \kmers from each sequence of a collection for later comparison. 